\documentclass[fleqn,usenatbib]{mnras}

\usepackage{newtxtext,newtxmath,cuted,lipsum}
\usepackage[T1]{fontenc} 
\usepackage{graphicx}	
\usepackage{amsmath}	
\usepackage{amssymb}	
\usepackage{amsfonts}
\usepackage{enumerate} 
\usepackage{colordvi} 
\usepackage{bm}
\usepackage{epsfig}
\usepackage{float}
\usepackage{verbatim}
\usepackage{subfig}
\usepackage[dvipsnames]{xcolor}
\usepackage{multirow,multicol}

\usepackage[colorinlistoftodos]{todonotes}
\usepackage{hyperref}
\usepackage{ulem}

\usepackage[colorinlistoftodos]{todonotes}
\usepackage{hyperref}

\newcommand{\bfk}{\mbox{\boldmath$k$}}
\newcommand{\bfp}{\mbox{\boldmath$p$}}

\newcommand{\B}[1]{\textcolor{black}{#1}} 
\definecolor{darkgreen}{cmyk}{0.85,0.2,1.00,0.05}

\usepackage{tikz}
\usetikzlibrary{shapes.geometric, arrows}

\tikzstyle{start} = [trapezium, trapezium left angle=70, trapezium right angle=110, minimum width=0.5cm, minimum height=0.5cm, text centered, draw=black, fill=yellow!80]

\tikzstyle{start_mg} = [trapezium, trapezium left angle=70, trapezium right angle=110, minimum width=0.5cm, minimum height=0.5cm, text centered, draw=black, fill=orange!50]

\tikzstyle{start_plus} = = [rectangle,  minimum width=0.5cm, minimum height=0.5cm,text centered]

\tikzstyle{startstop_scol} = [rectangle,  minimum width=3cm, minimum height=1cm,text centered, draw=black, fill=cyan!30]

\tikzstyle{startstop_halod} = [rectangle,  minimum width=3cm, minimum height=1cm,text centered, draw=black, fill=red!30]

\tikzstyle{startstop_halod2} = [rectangle,  minimum width=0.5cm, minimum height=0.5cm,text centered, draw=black, fill=red!30]

\tikzstyle{startstop_copter} = [rectangle,  minimum width=1.5cm, minimum height=0.8cm,text centered, draw=black, fill=green!30]

\tikzstyle{startstop_halod3} = [rectangle,  minimum width=1.5cm, minimum height=0.8cm,text centered, draw=black, fill=brown!80]

\tikzstyle{startstop_react} = [circle,  minimum width=1.5cm, minimum height=1.5cm,text centered, draw=black, fill=red!30]

\tikzstyle{io_mg} = [trapezium, trapezium left angle=70, trapezium right angle=110, minimum width=0.5cm, minimum height=0.5cm, text centered, draw=black, fill=orange!90]
\tikzstyle{io_halo} = [trapezium, trapezium left angle=70, trapezium right angle=110, minimum width=0.5cm, minimum height=0.5cm, text centered, draw=black, fill=red!90]

\tikzstyle{io_nl} = [trapezium, trapezium left angle=70, trapezium right angle=110, minimum width=2cm, minimum height=1cm, text centered, draw=black, fill=green!80]

\tikzstyle{io_pseudo} = [trapezium, trapezium left angle=70, trapezium right angle=110, minimum width=2cm, minimum height=1cm, text centered, draw=black, fill=gray!40]

\tikzstyle{connector} = [circle, minimum width=0.2cm, minimum height=0.2cm,text centered, draw=black, fill=black]

\tikzstyle{arrow} = [thick,->,>=stealth]

\title[On the road to per-cent accuracy IV:]{On the road to percent accuracy IV: {\tt ReACT} -- computing the non-linear power spectrum beyond $\Lambda$CDM}  

\author[B. Bose et al.]{Benjamin Bose$^{1}$\thanks{E-mail:benjamin.bose@unige.ch},
Matteo Cataneo$^{2}$,
Tilman Tr{\"o}ster$^{2}$,
Qianli Xia$^{2}$,
\newauthor
Catherine Heymans$^{2,3}$,
Lucas Lombriser$^{1}$
\\
$^{1}$D\'epartement de Physique Th\'eorique, Universit\'e de Gen\`eve, 24 quai Ernest Ansermet, 1211 Gen\`eve 4, Switzerland \\
$^{2}$Institute for Astronomy, University of Edinburgh, Royal Observatory, Blackford Hill, Edinburgh, EH9 3HJ, U.K. \\
$^{3}$German Centre for Cosmological Lensing (GCCL), Astronomisches Institut, Ruhr-Universit{\"a}t Bochum, Universit{\"a}tsstr. 150,
44801 Bochum, Germany.\\
}

\date{Accepted XXX. Received YYY; in original form ZZZ}

\pubyear{2019}

\begin{document}
\label{firstpage}
\pagerange{\pageref{firstpage}--\pageref{lastpage}}
\maketitle

\begin{abstract}
 To effectively exploit large-scale structure surveys, we depend on accurate and reliable predictions of non-linear cosmological structure formation. Tools for efficient and comprehensive computational modelling are therefore essential to perform cosmological parameter inference analyses. We present the public software package {\tt ReACT}, demonstrating its capability for the fast and accurate calculation of non-linear power spectra from non-standard physics.   We showcase {\tt ReACT} through a series of \B{ analyses} on the DGP and $f(R)$ gravity models, adopting LSST-like cosmic shear power spectra.  Accurate non-linear modelling with {\tt ReACT} \B{has the potential to} more than double LSST's constraining power on the $f(R)$ parameter, in contrast to an analysis that is limited to the quasi-linear regime.   We find that  {\tt ReACT} is sufficiently robust for the inference of consistent constraints on theories beyond $\Lambda$CDM for current and ongoing surveys. With further improvement, particularly in terms of the accuracy of the non-linear $\Lambda$CDM power spectrum, {\tt ReACT} can, in principle, meet the accuracy requirements for future surveys such as Euclid and LSST.
\end{abstract}

\begin{keywords}
cosmology: theory -- large-scale structure of the Universe -- methods: analytical
\end{keywords}

\section{Introduction}
The standard model of cosmology, based on general relativity plus a cosmological constant ($\Lambda$) and cold dark matter (CDM), has been extraordinarily successful in reproducing our cosmological observations such as the cosmic microwave background  \citep[CMB,][]{Planck:2015xua} and the large-scale structure of the universe \citep[LSS,][]{Anderson:2012sa,Song:2015oza,Beutler:2016arn}. The model relies on only two fundamental assumptions: that general relativity holds on all physical scales and that the Universe is homogeneous and isotropic. On the other hand, the so called `dark' components, $\Lambda$ and CDM, account for $95\%$ of the matter-energy content of the Universe today.

Further, several mild tensions in cosmological parameters between late-time measurements and the CMB have been uncovered. In particular, there is a tension in the value of the Hubble parameter today, $H_0$, \citep{Efstathiou:2013via,Zhang:2017aqn,Riess:2019cxk,Wong:2019kwg,Lombriser:2019ahl,Pesce:2020xfe}  and in the amplitude of density fluctuations, $\sigma_8$, \citep{Heymans:2013fya,Hildebrandt:2016iqg,Abbott:2017wau} from direct measurement and inferred from extrapolating from the best fit CMB data \citep{Planck:2015xua} \citep[also see][]{Lin:2017bhs}.

Motivated by these issues, probing the nature of dark matter and dark energy, as well as testing alternatives to $\Lambda$CDM,  is the main focus of modern cosmology. 
In particular, a plethora of exotic dark energy and modified gravity models have been proposed over the past couple of decades  \citep[for reviews see][]{Copeland:2006wr,Clifton:2011jh,Joyce:2016vqv,Koyama:2018som}. But any viable alternative to the concordance model must pass all Solar System tests, match all cosmological data equally well and moreover, not modify the speed of gravitational wave propagation \citep{Lombriser:2015sxa,Monitor:2017mdv,Lombriser:2016yzn,Creminelli:2017sry,Ezquiaga:2017ekz,Baker:2017hug,Sakstein:2017xjx,Battye:2018ssx,deRham:2018red,Creminelli:2018xsv}. This places very tight constraints on modifications to $\Lambda$CDM in the regimes of these experiments. 

One regime that still remains largely open to signals of modified gravity or dark energy is the LSS of the universe, in particular, the non-linear, small cosmological scales. It is in this regime that cosmological modifications of gravity are expected to give clear signatures as they transition to recover general relativity at Solar-System scales. Many modified gravity models generically realise such a transition through screening mechanisms \citep{Vainshtein:1972sx,Khoury:2003rn,Babichev:2009ee,Hinterbichler:2010es}. With current and future LSS surveys such as LSST\footnote{ The Vera C. Rubin Observatory Legacy Survey of Space and Time: \url{https://www.lsst.org/}} \citep{Chang:2013xja}, KiDS \footnote{The Kilo Degree Survey: \url{http://kids.strw.leidenuniv.nl/}} \citep{Kuijken:2015vca}, DES \footnote{The Dark Energy Survey: \url{https://www.darkenergysurvey.org/}} \citep{Albrecht:2006um}, DESI\footnote{The Dark Energy Spectroscopic Instrument: \url{http://desi.lbl.gov/}} \citep{Aghamousa:2016zmz}, HSC\footnote{\url{Hyper Suprime-Cam: https://hsc.mtk.nao.ac.jp/ssp/survey/}} \citep{Aihara:2017paw} and Euclid \footnote{Euclid: \url{www.euclid-ec.org}} \citep{Laureijs:2011gra}, our ability to precisely measure these scales has become unprecedented and will further improve over the next decade. With this new wealth of high-precision data in the non-linear regime of cosmic structure formation, the challenge has become the accurate and efficient modelling of our observables for our theories.

In this paper, we present {\tt ReACT} \footnote{Download {\tt ReACT}: \url{https://github.com/nebblu/ReACT}}, a fast and reliable code that employs the approach of \cite{Cataneo:2018cic} to accurately compute general modifications to the non-linear $\Lambda$CDM matter power spectra for theories beyond $\Lambda$CDM. The efficiency of {\tt ReACT} enables its implementation in statistical parameter inference pipelines, as typically employed by LSS surveys. We demonstrate the reliable performance of the code in a set of Markov Chain Monte Carlo (MCMC) analyses on mock LSS data sets that aim to represent future survey cosmic shear measurements.
These furthermore double as preliminary analyses that investigate the importance of modelling non-linear modifications of gravity in cosmological and gravitational parameter estimation.

This paper is organised as follows: In section~\ref{sec:theory} we present the theoretical framework used to compute general modifications to $\Lambda$CDM non-linear power spectra. In section~\ref{sec:code} we present the {\tt ReACT} code, outlining its structure, performance and a useful implementation into the well established {\tt CosmoSIS} \citep{Zuntz:2014csq} framework. We present our MCMC analyses in section~\ref{sec:analysis}. We summarise our results and conclude in section~\ref{sec:summary}.


\section{The framework: halo model reactions }\label{sec:theory}

In order to effectively exploit the high-precision data supplied by the new generation of galaxy surveys, which may reveal new physics, we rely on accurate and reliable computational tools for modelling the non-linear cosmic structure formation.
Here, we are concerned with computing the non-linear matter power spectrum in a model-independent framework. In \cite{Cataneo:2018cic}, the reaction approach was introduced, providing such a framework. Its prescription for the non-linear matter power spectrum in the desired theory of dark energy or gravity, $P_{\rm NL}$, involves the two quantities explicit in the following equation  
\begin{equation}
    P_{\rm NL}(k,z) = \mathcal{R}(k,z) P^{\rm pseudo}_{\rm NL} (k,z). \label{eq:nlps}
\end{equation}
Here, $\mathcal{R}(k,z)$ is the halo model reaction and $P^{\rm pseudo}_{\rm NL}(k,z)$ is the pseudo-matter $\Lambda$CDM  power spectrum. Note that all power spectra without a superscript will be assumed to be the beyond-$\Lambda$CDM spectra.

\subsection{Pseudo power spectrum}
The pseudo cosmology is defined as a universe with $\Lambda$CDM physics but where the initial conditions are adjusted so that the linear clustering matches that of the target beyond-$\Lambda$CDM cosmology, i.e. 
\begin{equation}
    P_{\rm L}^{\rm pseudo}(k,z) = P_{\rm L}(k,z), 
\end{equation}
where $P_{\rm L}(k,z)$ is the linear power spectrum in the theory of interest, be it modified gravity or having an evolving dark energy component (or both). \B{ In practice this  is done by rescaling the modified linear power spectrum at the target redshift to the initial redshift using the linear $\Lambda$CDM growth factor.}

Modelling the non-linear pseudo power spectrum accurately is essential in obtaining percent-level accuracy in the target spectrum $P_{\rm NL}(k,z)$.  \cite{Giblin:2019iit} propose the development of an emulator for this quantity. In this work, we will use the halo-model based formula described in \cite{Mead:2015yca, Mead:2016zqy} to give predictions for $ P^{\rm pseudo}_{\rm NL} (k,z)$, which is accurate at the $5\%$-level for $k<10h/{\rm Mpc}$. We note that this level of accuracy is the key limiting factor of our predictions for the target spectrum,  $ P_{\rm NL} (k,z)$, at scales $k\leq 5h/{\rm Mpc}$. At smaller scales, the inaccuracy of the reaction, $\mathcal{R}$, becomes of the same order. Note that the modified linear spectrum, $P_{\rm L}(k,z)$ is computed using {\tt ReACT}. 

\B{To obtain $ P^{\rm pseudo}_{\rm NL} (k,z)$, we provide the target modified $P_{\rm L}(k,z)$ as input for the publicly available {\tt HMCode}, which computes the halo model function of \cite{Mead:2015yca, Mead:2016zqy}.} Further, we emphasise that the choice of the \cite{Mead:2015yca} formula for $P_{\rm NL}^{\rm pseudo}$ is an implementation detail and our framework will work for any prescription for the non-linear $P(k)$ in $\Lambda$CDM as long as one can specify the linear clustering to the chosen framework. \B{In other words, one can model this spectrum using any standard $\Lambda$CDM method that can be informed of the non-standard changes in the shape of the beyond-$\Lambda$CDM linear power spectrum. For more details on the pseudo spectrum we refer the reader to section 3.3 of \cite{Cataneo:2018cic} and \cite{Giblin:2019iit}.}

\subsection{The reaction}
The quantities {\tt ReACT} computes are both $P_{\rm L}(k,z)$ and the halo-model reaction  $\mathcal{R}(k,z)$, that is specified by \citep{Cataneo:2018cic} 

\begin{equation}
\mathcal{R}(k,z) =  \frac{[(1-\mathcal{E}(z))e^{-k/k_\star(z)} + \mathcal{E}(z)] P_{\rm L}(k,z)  +  P_{\rm 1h}(k,z)}{P_{\rm hm}^{\rm pseudo}(k,z)}. \label{eq:reaction}
\end{equation}

\noindent It is an ansatz for the response of a $\Lambda$CDM spectrum to modified physics, based on the halo model and 1-loop perturbation theory (see \cite{Cooray:2002dia} and \cite{Bernardeau:2001qr} for respective reviews). The different components are given as

\begin{align}
  P_{\rm hm}^{\rm pseudo}(k,z) = &   P_{\rm L} (k,z) + P_{\rm 1h}^{\rm pseudo}(k,z), \\ 
  \mathcal{E}(z) =& \lim_{k\rightarrow 0} \frac{ P_{\rm 1h}^{\rm }(k,z)}{ P_{\rm 1h}^{\rm pseudo}(k,z)} , \label{mathcale} \\ 
   k_{\rm \star}(z) = & - \bar{k} \left(\ln \left[ 
    \frac{A(\bar{k},z)}{P_{\rm L}(\bar{k},z)} - \mathcal{E}(z) \right] - \ln\left[1-\mathcal{E}(z) \right]\right)^{-1}, \label{kstar}
\end{align}
where 
\begin{equation}
    A(k,z) =  \frac{P_{\rm 1-loop}(k,z)+ P_{\rm 1h}(k,z)}
    {P^{\rm pseudo}_{\rm 1-loop}(k,z)+ P_{\rm 1h}^{\rm pseudo}(k,z)}  P_{\rm hm}^{\rm pseudo}(k,z) -  P_{\rm 1h}(k,z).
\end{equation}
$P_{\rm 1h}(k,z)$ and $P_{\rm 1h}^{\rm pseudo} (k,z)$ are the 1-halo terms predicted by the halo model with and without non-linear modifications to $\Lambda$CDM respectively. Remember that the pseudo 1-halo term requires linear clustering to be described in the modified theory as per definition of the `pseudo' cosmology. These quantities are described in more detail in appendix \ref{app:sphercol}. 

$P_{\rm 1-loop}(k,z)$ and $P_{\rm 1-loop}^{\rm pseudo} (k,z)$ are the 1-loop power spectra predicted by standard perturbation theory with and without non-linear modifications to $\Lambda$CDM. In effect, the pseudo 1-loop spectrum is well described by the `un-screened' spectrum in modified gravity. This boils down to setting the 2nd and 3rd order Poisson equation modifications $\gamma_2 = \gamma_3 = 0$ (see appendix \ref{app:modgpt} for more details). Note that this approximation is affected by similar inaccuracies characterising the `screened' 1-loop spectrum, therefore it is more suitable for fractional quantities like the reaction.

For the limit in equation~\eqref{mathcale} we take $k=0.01h/{\rm Mpc}$ which is sufficient for all redshifts up to $z=0$. For the scale at which we compute $k_\star$, we choose $\bar{k} = 0.06h/{\rm Mpc}$ following \cite{Cataneo:2018cic}. This scale is chosen such that the 1-loop perturbative predictions are sufficiently accurate at all redshifts considered. 

We now move on to describe the code {\tt ReACT} which produces predictions for equation~\eqref{eq:reaction} under some specification of the background Hubble function $H(z)$, the linear, 2nd and 3rd order Poisson equation modifications $\mu,\gamma_2,\gamma_3$ \citep{Bose:2016qun} as well as the spherical collapse modification $\mathcal{F}$ \citep[see for example][]{Lombriser:2013eza}. This parametrisation is explicitly specified, for the theories we consider, in appendix \ref{app:parametrisations}.  


\section{The code: {\tt ReACT}}\label{sec:code}
In this section we give some details about {\tt ReACT}. The code is written in {\tt C++} and is based on the 1-loop perturbation theory code {\tt MG-Copter} \citep{Carlson:2009it,Bose:2016qun}. {\tt MG-Copter} provides a means of numerically calculating the 1-loop matter power spectrum using the algorithm described in \cite{Taruya:2016jdt}, which is applicable for general theories of gravity and dark energy \citep{Bose:2016qun,Bose:2017jjx}. We have optimised this code to compute the 1-loop matter spectrum much faster than the original version presented in \cite{Bose:2016qun}. The optimisation also corrects the sampling-dependent accuracy described in \cite{Bose:2016qun}. 

We have added two new libraries to {\tt MG-Copter}. The first, {\tt SCOL}, handles both the spherical collapse and virial theorem computations outlined in appendix \ref{app:sphercol}, which are needed for the halo-model spectra computations. To do this, we make use of the non-linear differential equation solver package, {\tt SUNDIALS}\footnote{\url{https://computation.llnl.gov/projects/sundials}}. The equations governing spherical collapse can also be edited fairly easily \citep[for example if one wishes to employ ellipsoidal collapse, see][]{Sheth:1999su}.

The second library, {\tt HALO}, contains all the relevant quantities needed to compute $\mathcal{R}(k,z)$. In particular, the following key quantities are explicit functions within this library; the halo mass function $n_{\rm vir}(M_{\rm vir},z)
$, the virial concentration $c_{\rm vir}(M_{\rm vir}, z)$ and the Fourier transform of the halo density profile $u(k,M_{\rm vir},z)$. This makes it easy for the user to edit the form of these quantities for both pseudo and fully modified cosmologies. The default forms for these, and the ones used in the proceeding analyses, are the Sheth-Torman mass function \citep{Sheth:1999mn,Sheth:2001dp}, a power law virial concentration \citep[see for example][]{Bullock:1999he} and the halo density profile described in \cite{Navarro:1996gj}.

Together with the perturbation theory library, {\tt SPT}, used to calculate equation~\eqref{kstar}, one can compute the reaction, equation~\eqref{eq:reaction}, for a given $\Lambda$CDM input power spectrum produced by a Boltzmann equation solver such as {\tt CAMB} \citep{Lewis:2002ah} \footnote{\url{http://camb.info}} and a specified model of gravity or dark energy. 

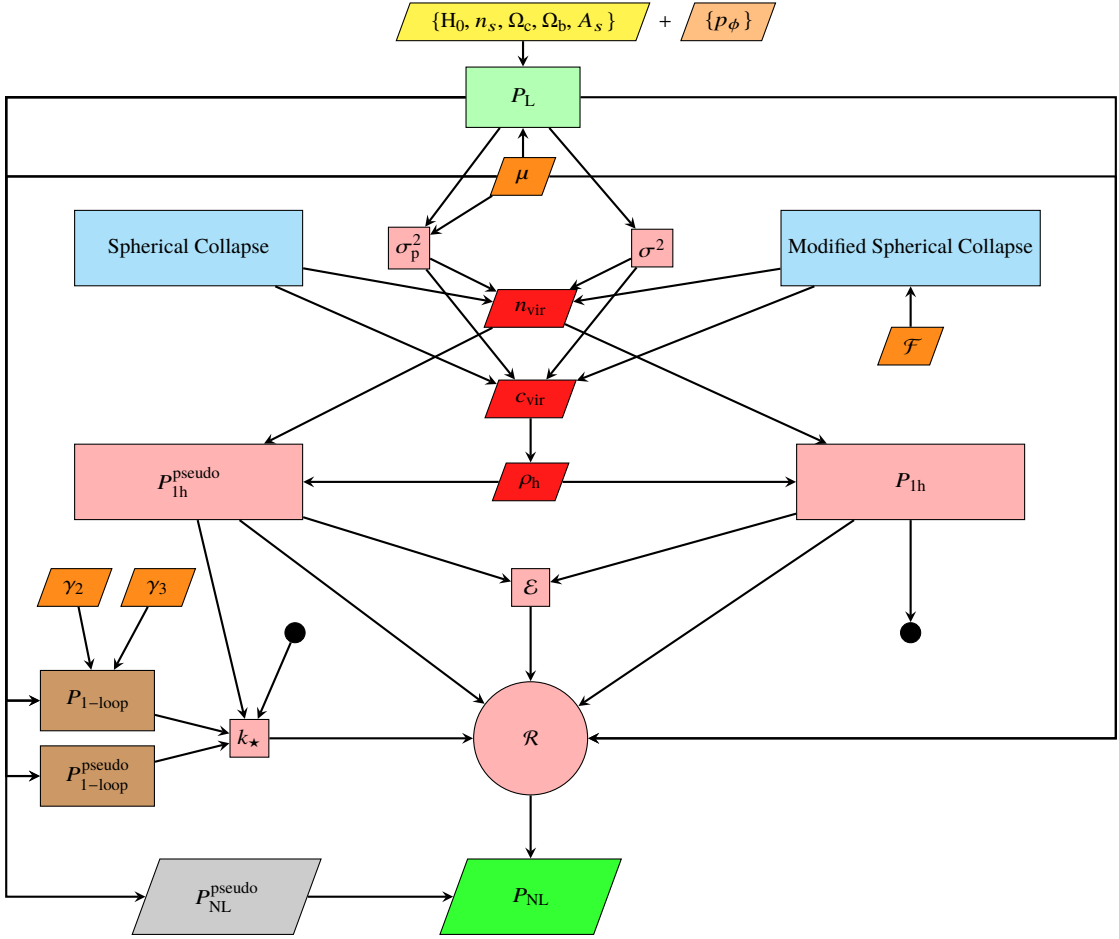
\begin{figure*}
\centering
\begin{tikzpicture}[node distance=2cm]
\node (pseudo1h) [startstop_halod,xshift=1cm] {$P_{\rm 1h}^{\rm pseudo}$};
\node (real1h) [startstop_halod, right of=pseudo1h,xshift=7.5cm] {$P_{\rm 1h}$};
\node (sc) [startstop_scol, above of=pseudo1h,xshift=0cm,yshift=1.1cm] {Spherical Collapse};
\node (mgsc) [startstop_scol, above of=real1h,xshift=0cm,yshift=1.1cm] {Modified Spherical Collapse};
\node (mgfunc) [io_mg, below of=mgsc, xshift=0cm,yshift=0.7cm] {$\mathcal{F}$};
\node (conn) [connector, below of=real1h,yshift=-0cm] {};
\node (rho_h) [io_halo ,left of=pseudo1h,xshift=6.5cm,yshift=0cm] {$\rho_{\rm h}$};
\node (c_vir) [io_halo, above of=rho_h,yshift=-0.9cm] {$c_{\rm vir}$};
\node (n_vir) [io_halo, above of=c_vir,yshift=-0.8cm,xshift=-0.cm] {$n_{\rm vir}$};
\node (sig2) [startstop_halod2 ,above of=c_vir,yshift=0cm,xshift=1.6cm] {$\sigma^2$};
\node (sig2b) [startstop_halod2 ,above of=c_vir,yshift=0cm,xshift=-1.6cm] {$\sigma^2_{\rm p}$};
\node (plpseudo) [startstop_copter,above of=sig2b,xshift=1.5cm,yshift=-0cm] {$P_{\rm L}$};
\node (mu) [io_mg ,below of=plpseudo,yshift=0.95cm,xshift=0cm] {$\mu$};
\node (bige) [startstop_halod2 , below of=rho_h,yshift=0.6cm,xshift=0cm] {$\mathcal{E}$};
\node (cosmology) [start ,above of=plpseudo, yshift=-1cm] {${\rm \{ H_0},n_s,\Omega_{\rm c},\Omega_{\rm b}, A_s\}$};
\node (plus) [start_plus ,right of=cosmology, xshift=-0.13cm] {$+$};
\node (modification) [start_mg ,right of=cosmology, xshift=0.7cm] {$\{ p_{\rm \phi}\}$};
\draw [arrow] (real1h) -- (conn);
\draw [arrow] (cosmology) -- (plpseudo);
\draw [arrow] (mgfunc) -- (mgsc);
\draw [arrow] (c_vir) -- (rho_h);
\draw [arrow] (sig2) -- (c_vir);
\draw [arrow] (sig2b) -- (c_vir);
\draw [arrow] (mu) -- (plpseudo);
\draw [arrow] (mgsc) -- (c_vir);
\draw [arrow] (sc) -- (c_vir);
\draw [arrow] (mgsc) -- (n_vir);
\draw [arrow] (sc) -- (n_vir);
\draw [arrow] (plpseudo) -- (sig2b);
\draw [arrow] (plpseudo) -- (sig2);
\draw [arrow] (mu) -- (sig2b);
\draw [arrow] (n_vir) -- (real1h);
\draw [arrow] (n_vir) -- (pseudo1h);
\draw [arrow] (sig2) -- (n_vir);
\draw [arrow] (sig2b) -- (n_vir);
\draw [arrow] (rho_h) -- (pseudo1h);
\draw [arrow] (rho_h) -- (real1h);
\draw [arrow] (pseudo1h) -- (bige);
\draw [arrow] (real1h) -- (bige);
\node (kstar) [startstop_halod2 ,below of=pseudo1h,yshift=-1.4cm,xshift=0.8cm] {$k_
\star$};
\node (conn2) [connector, above of=kstar,yshift=-0.6cm,xshift=0.6cm] {};
\node (1loopp) [startstop_halod3 ,left of=kstar,yshift=-0.5cm,xshift=0cm] {$P_{\rm 1-loop}^{\rm pseudo}$};
\node (1loopr) [startstop_halod3 ,left of=kstar,yshift=0.5cm,xshift=0cm] {$P_{\rm 1-loop}$};
\node (gamma2) [io_mg ,above of=1loopr,xshift=-0.3cm,yshift=-0.5cm] {$\gamma_2$};
\node (gamma3) [io_mg ,above of=1loopr,xshift=0.8cm,yshift=-0.5cm] {$\gamma_3$};
\draw [arrow] (pseudo1h) -- (kstar);
\draw [arrow] (conn2) -- (kstar);
\draw [arrow] (gamma2) -- (1loopr);
\draw [arrow] (gamma3) -- (1loopr);
\draw [arrow] (1loopr) -- (kstar);
\draw [arrow] (1loopp) -- (kstar);
\draw [arrow] (plpseudo) --  node[anchor=west, above=2pt ] {} ++(-6.8,0) |- (1loopr);
\draw [arrow] (plpseudo) --  node[anchor=west, above=2pt ] {} ++(-6.8,0) |- (1loopp);
\draw [arrow] (mu) --  node[anchor=west, above=2pt ] {} ++(-6.8,0) |- (1loopp);
\draw [arrow] (mu) --  node[anchor=west, above=2pt ] {} ++(-6.8,0) |- (1loopr);
\node (reaction) [startstop_react ,below of=bige,xshift=0cm] {$\mathcal{R}$};
\draw [arrow] (mu) --  node[anchor=west, above=2pt ] {} ++(7.8,0) |- (reaction);
\draw [arrow] (pseudo1h) -- (reaction);
\draw [arrow] (real1h) -- (reaction);
\draw [arrow] (kstar) -- (reaction);
\draw [arrow] (bige) -- (reaction);
\draw [arrow] (plpseudo) --  node[anchor=west, above=2pt ] {} ++(7.8,0) |- (reaction);
\node (pnl) [io_nl ,below of=reaction,yshift=-0.1cm] {$P_{\rm NL}$};
\node (pnlpseudo) [io_pseudo ,left of=pnl,xshift=-2cm] {$P_{\rm NL}^{\rm pseudo}$};
\draw [arrow] (reaction) -- (pnl);
\draw [arrow] (pnlpseudo) -- (pnl);
\draw [arrow] (plpseudo) --  node[anchor=west, above=2pt ] {} ++(-6.8,0) |- (pnlpseudo);

\end{tikzpicture}
 \caption[CONVERGENCE]{An overview of the computation of the non-linear power spectrum. The yellow trapezoid (top, centre) indicates the standard input cosmological parameter set in $\Lambda$CDM. These are specified to the Boltzmann solver which computes the unmodified linear power spectrum. The light orange trapezoid ($p_{\phi}$) indicates additional parameters that describe modifications to $\Lambda$CDM. These must also be specified, either within {\tt ReACT} or with the other cosmological parameters in a {\tt .ini}  file when using {\tt CosmoSIS}. Brown boxes ($P_{\rm 1-loop}^{\rm pseudo}$ and $P_{\rm 1-loop}$) indicate computations performed by standard {\tt MG-Copter} ({\tt SPT} and {\tt SpecialFunctions} libraries). The light blue boxes ((Modified) Spherical Collapse) indicate computations performed by the {\tt SCOL} library. The pink boxes ($\sigma_{\rm p}^2$, $\sigma^2$, $P^{\rm pseudo}_{\rm 1h}$, $P_{\rm 1h}$, $\mathcal{E}$, $k_\star$ and $\mathcal{R}$) indicate computations performed by the {\tt HALO} library. The red trapezoids ($n_{vir}$, $c_{vir}$ and  $\rho_h$) indicate key quantities in the halo model, all found explicitly in {\tt HALO}. The bright orange trapezoids ($\mu$, $\gamma_2$, $\gamma_3$ and $\mathcal{F}$) indicate quantities related to the modification to $\Lambda$CDM, found in {\tt SpecialFunctions}. These quantities are implemented such that they can be easily modified by the user and that, although not explicit in the flowchart, require $p_\phi$ as input. The black dot connects the $P_{\rm 1h}$ box to the $k_\star$ box. Also note that $\sigma_{\rm p}^2$ is the variance used in the $P_{\rm 1h}^{\rm pseudo}$ calculation.} 
\label{algorithm}
\end{figure*}

Figure~\ref{algorithm} presents a schematic of the computation with the quantities roughly colour coded according to the library they are part of. Note that the orange boxes indicate functions that need to be specified by the user for the model of gravity under consideration. Currently, the code includes the following presets with respective theory parameters: \cite{Hu:2007nk} $f(R)$ gravity $\{f_{R0} \}$ , the braneworld model DGP of \cite{Dvali:2000hr}  $\{ \Omega_{\rm rc} = 1/(4H_0^2r_c^2) \}$, wCDM  $\{w_0, w_a\}$ and $\Lambda$CDM. Note that for $f(R)$ and DGP we assume a $\Lambda$CDM background, but for wCDM one should also modify the Hubble function within {\tt ReACT}. Various evolving dark energy backgrounds have been included within the code. We direct the reader to appendix~\ref{app:parametrisations} for the details.

We provide an easy-to-use Python interface to {\tt ReACT}, called {\tt pyreact}. This interface has also been integrated into the {\tt CosmoSIS} \citep{Zuntz:2014csq} parameter inference framework.

We note that all components of the reaction (as well as the reaction itself) are computed by {\tt ReACT} to better than $0.5\%$ accuracy when compared with the computations of \cite{Cataneo:2018cic}, for all models and redshifts considered in that work ($f(R)$, DGP and wCDM at $z=0,1$). 

\section{The analysis: Impact of {\tt ReACT}}\label{sec:analysis}
We now proceed with both demonstrating the \B{speed and consistency} of {\tt ReACT} and provide some preliminary modified gravity forecasts. We do this by conducting two sets of parameter inference analyses employing an implementation of {\tt ReACT} in {\tt CosmoSIS}. For these analyses we generate \B{ mock data consisting of a $C_\ell^{ ij}$ data vector given by \cite{Hildebrandt:2016iqg}}
\begin{equation}
C_{l}^{ij} = \int_{0}^{\chi_{\rm H}} d \chi \frac{g_i(\chi) g_j(\chi) }{\chi^{2}} P_{\rm NL}\left(k=\frac{l}{\chi(z)}, z(\chi)\right) \, ,
\label{lensingconv}
\end{equation}
\B{where $P_{\rm NL}(k, z)$ is the matter power spectrum at redshift $z$ and the co-moving distance, $\chi$ is integrated from 0 to the horizon distance $\chi_H$. The weights are given by }
\begin{equation}
    g_i(\chi)=\frac{3 \Omega_m H_0^2 \chi}{2 c^2 a} \int_{\chi(z)}^{\chi_H} n_i(z) \frac{\chi(z')-\chi(z) }{\chi(z')}dz' \, , 
\end{equation}
\B{with $n_i(z)$ being the galaxy distribution in the ith bin (see equation~\ref{eq:galdist}). $P_{\rm NL}(k, z)$ is generated using equation~\eqref{eq:nlps}, with $ P^{\rm pseudo}_{\rm NL} (k,z)$ being generated using {\tt HMCode}.} 

\B{The mock data also consists of a Gaussian covariance matrix (see equation 2.9 of \cite{Barreira:2018jgd} for example). For the covariance matrix} we assume a stage IV-like survey such as LSST or Euclid, and so adopt the following specifications: sky fraction $f_{\rm sky} = 0.436$, number density of galaxies per arcminute squared $n = 30\, \mathrm{arcmin}^{-2}$ and shape noise parameter $\sigma_e = 0.3$ \citep[see][]{Zhan:2017uwu,Amendola:2012ys}. Further, we take 3 tomographic bins in the redshift range $0 \leq z \leq 2$ with bin edges defined by $[0.00, 0.48, 0.81, 2.00]$\footnote{These bin widths are chosen to keep the number density of objects roughly the same over all bins.}, and unless otherwise stated, we use 19 $\ell$ bins in the range $[20,3000]$.  The galaxy distribution follows the following relation \citep{Smail:1994sx,Chang:2013xja}
\begin{equation}
    n(z) = z^2 \exp \left[-\frac{z}{0.24} \right] \, . \label{eq:galdist}
\end{equation}

We perform two sets of analyses. \B{The first set applies predictions for the shear spectrum, equation~\ref{lensingconv}, using the non-linear matter power spectrum predicted using equation~\eqref{eq:nlps}} to mock data generated with the $\Lambda$CDM cosmology described in table~\ref{tab:cos}. \B{In this case we have $\mathcal{R}=1$ and a $\Lambda$CDM $P_{\rm L}(k;z)$. To this data we fit predictions within} $f(R)$ and DGP separately, with $f_{R0}$ kept free and then $\Omega_{rc}$ kept free, \B{each time} setting the appropriate parametrisation within {\tt ReACT}  (see appendix~\ref{app:parametrisations}) in each instance. This aims to give a rough forecast of the level of constraining power future surveys will have on these parameters. 

\begin{table}
\centering
\caption{Cosmological parameter values and priors for the mock data used in all analyses. We note that the derived parameters $\sigma_8= 0.844$, $\Omega_m=0.314$ and $S_8 = 0.863$ for this cosmology. We also show the priors and starting values for the modified gravity parameters used in the $\Lambda$CDM mock data analyses. }
\begin{tabular}{| c | c | }
\hline  
 Parameter & Mock value and flat prior bounds  \\
 \hline
 $H_0$ & [50, 67.3, 80] \\ \hline 
  $n_s$ & [0.9, 0.966, 1.05] \\ \hline 
 $\Omega_c$ & [0.15, 0.265, 0.36] \\ \hline 
 $\Omega_b$ & [0.01, 0.049, 0.13]  \\ \hline 
 $10^9 A_s$ & [1.65, 2.2, 2.55]  \\ \hline \hline 
  Log$[{f}_{R0}]$ & [-8, -7, -5]  \\ \hline 
  $\Omega_{\rm rc} $ & [0, 0.1, 1.5]  \\ \hline 
\end{tabular}
\label{tab:cos}
\end{table}

The second type of analyses will use the same covariance matrix as in the $\Lambda$CDM mock data, but the data vector will be produced using {\tt ReACT} + {\tt HMCode} with $f(R)$ as a preset. We choose the fiducial parameter value Log$[f_{\rm R0}]=-5$ (F5). We will fit this data for different scale cuts, the aim being to get an idea of how constraining the non-linear scales are in the case of an $f(R)$ cosmology. We also fit this data with the pseudo power spectrum (an $\mathcal{R}=1$ model) to investigate the impact of omitting the non-linear corrections described by equation~\eqref{eq:reaction}.  

\B{We note that these analyses are primarily meant to demonstrate the code and not meant to illustrate the accuracy of the framework, which in fact has already been studied in \cite{Cataneo:2018cic} using full N-body simulations. We further investigate the accuracy of the reaction approach in appendix~\ref{app:accuracy}. Secondary is the investigation of what type of constraints on gravity one can expect from a cosmic shear analysis using various scale cuts.}

For all analyses we use the {\tt emcee} \citep{ForemanMackey:2012ig} sampler module in {\tt CosmoSIS} with 32 walkers. We sample over the standard cosmological parameters $\{H_0, n_s, \Omega_c, \Omega_b, A_s \}$ as well as the modified gravity parameters $f_{R0}$ and $\Omega_{rc}$ for their respective analyses. $\sigma_8$, $\Omega_m$ and $S_8 = \sigma_8 \sqrt{\Omega_m/0.3}$ distributions are shown as derived parameters. In all analyses we use a minimum of 200,000 samples, and continue to sample until the distributions exhibit convergence. Rather than give hardware specific estimates on the CPU time taken for these analyses, we observe that analyses including the reaction computation are $\sim 5$ times slower than the same computation in $\Lambda$CDM.  The bottle-neck computation is the spherical collapse computation which needs to be done multiple times over a reasonable halo  mass range, as in modified gravity the collapse density will depend on halo mass. This can be optimised with a careful choice of differential equation solver and root finder. To a lesser extent, the 1-loop perturbation theory computations are also fairly slow. This computation is done numerically as described in \cite{Taruya:2016jdt} and \cite{Bose:2016qun}. Again, this may be optimised with a better choice of differential equation solver and/or tuning of the accuracy demands of the solver and 1-loop integrals. 

\subsection{$\Lambda$CDM mock data}
In this section we show results for analyses with different scale cuts: $\ell_{\rm max} = 3000,$ $\ell_{\rm max} = 1500$ and $\ell_{\rm max} = 500$ using $19$, $15$ and $8$ $\ell$-bins respectively, where $\ell_{\rm max}$ is the maximum multipole included in the cosmic shear analysis. Table~\ref{fittable} shows the upper $68\%$ and $99.7\%$ confidence limits on the marginalised modified gravity parameters for the various analyses discussed in this section. We warn the reader that these conclusions may depend on inaccuracies in the reaction and {\tt HMCode} \citep[see appendix~\ref{app:accuracy} and figure 1 of][]{Cataneo:2018cic}.

\begin{figure*}
\centering
  \includegraphics[width=\textwidth,height=\textwidth]{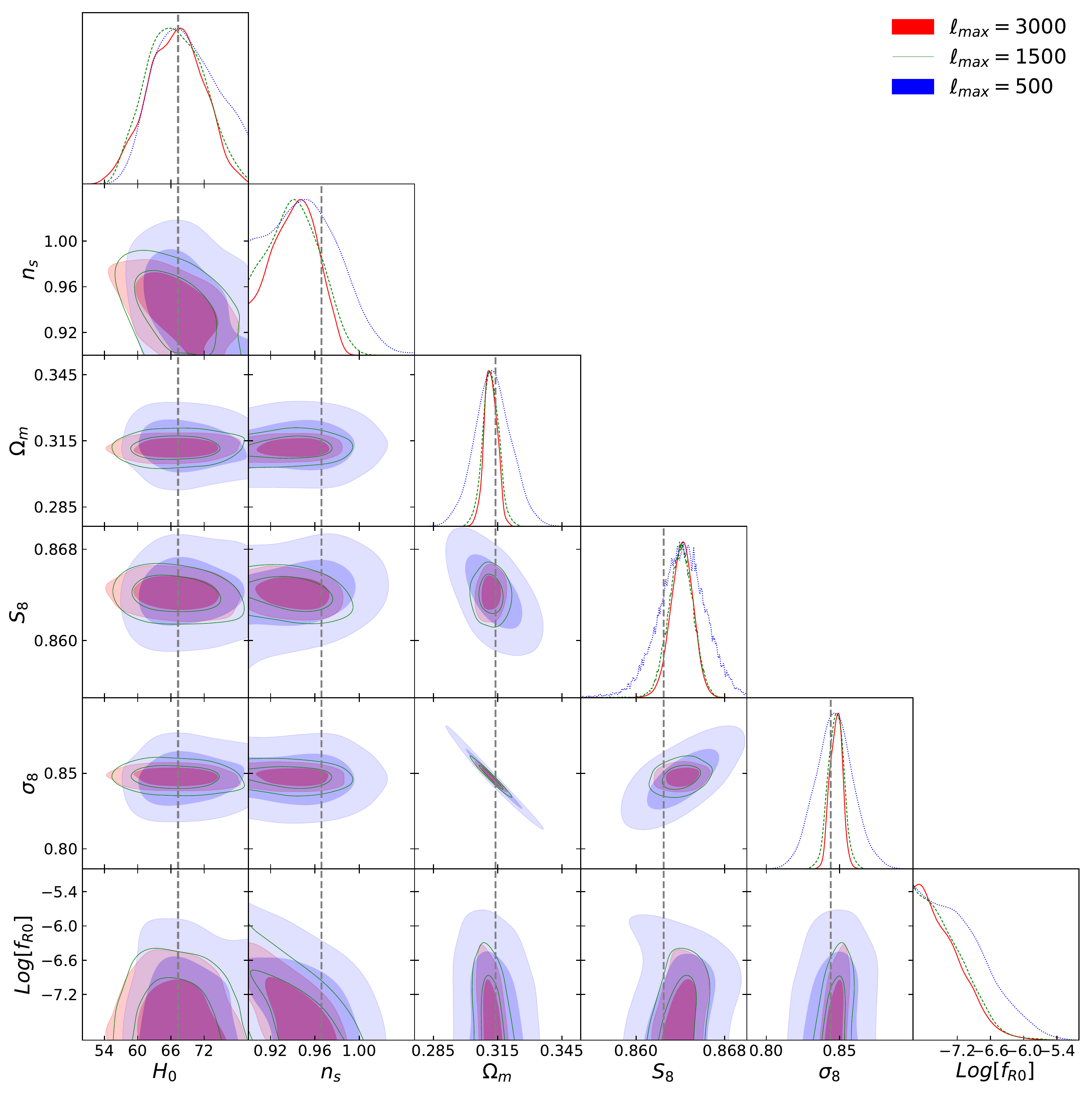}  \\  
  \caption[CONVERGENCE]{MCMC results for an $f(R)$ data analysis of a $\Lambda$CDM mock data vector within an LSST-like survey, using $\ell_{\rm max} = 3000,1500,500$ in red (filled contours, solid), green (unfilled contours, dashed) and blue (filled contours, dotted) respectively. Modelling of $C_\ell$ uses {\tt ReACT} + {\tt HMCode} (see equation~\ref{eq:nlps}). The mock data fiducial cosmological values are marked as dashed lines.}
\label{figurefrellmax}
\end{figure*}

Figure~\ref{figurefrellmax} shows  the two-dimensional marginalised posterior distributions when we fit the $\Lambda$CDM-mock data using an $f(R)$ model. Here we find negligible differences in the marginalised constraints on $f_{\rm R0}$ between the $\ell_{\rm max} = 3000$ and $1500$ analyses. This is probably driven by the current accuracy of the {\tt HMCode}-pseudo spectrum and the reaction. With more accurate prescriptions, one should be able to discern between these cases \citep[see for example][]{Giocoli:2018gqh}. In these cases one can rule out ${\rm Log}[f_{ \rm R0}]=-6$ at roughly the $99.7\%$ confidence level. 

On the other hand, the constraints are significantly degraded when using $\ell_{\rm max} = 500$, which only include scales that are within the quasi non-linear regime. In this case we can only rule out ${\rm Log} [f_{\rm R0}] = -5.5$ at $99.7\%$ confidence, which is close to existing constraints \citep{Pratten:2016ojf,Harnois-Deraps:2015ula}. We have also fitted this data with the pseudo power spectrum, i.e. setting $\mathcal{R}=1$, for $\ell_{\rm max}=3000$, and observe no significant differences in the parameter distributions when compared with the fully non-linear modelling i.e. with $\mathcal{R}$ included. 

\begin{figure*}
\centering
  \includegraphics[width=\textwidth,height=\textwidth]{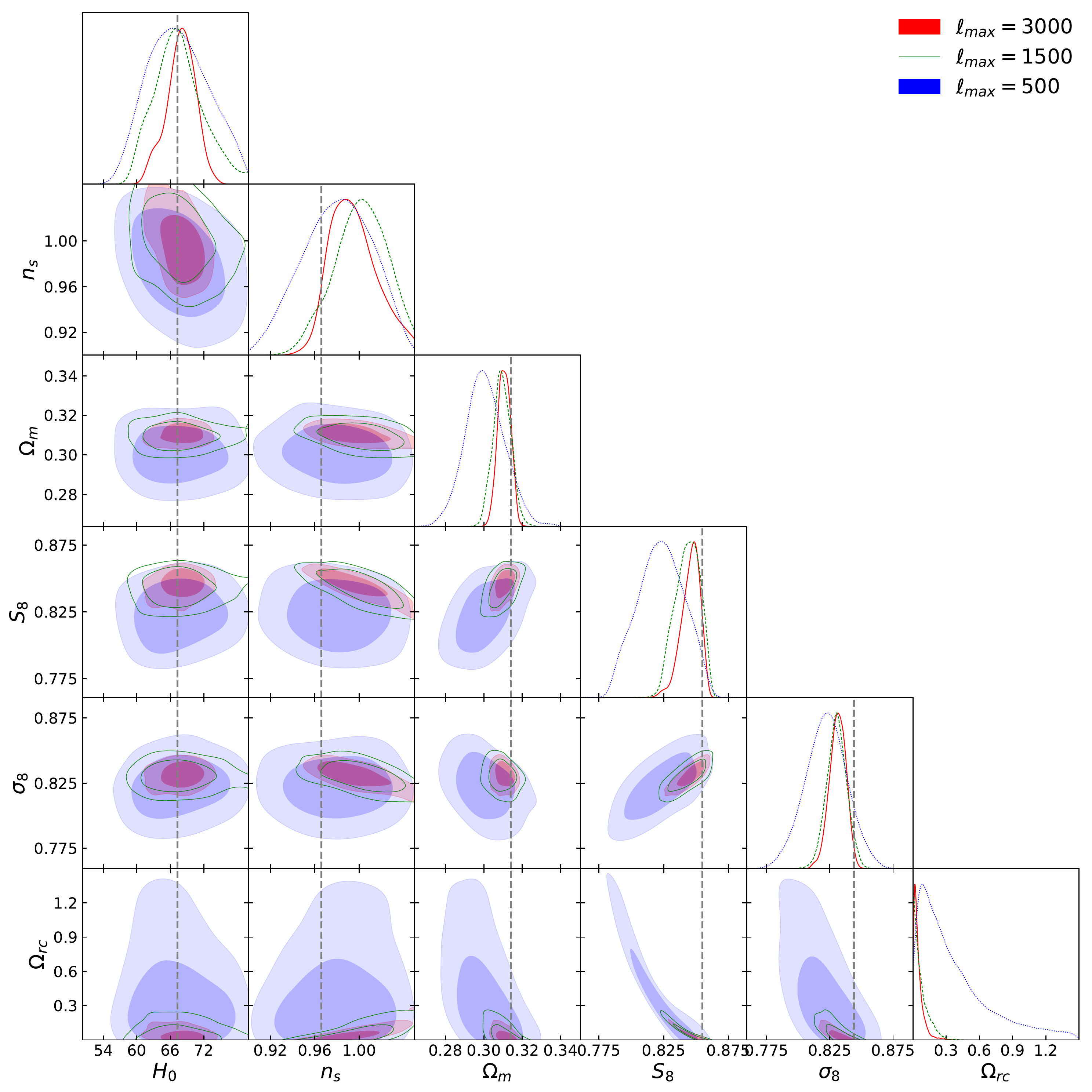}  \\  
  \caption[CONVERGENCE]{MCMC results for a DGP data analysis of a $\Lambda$CDM mock data vector within an LSST-like survey, using $\ell_{\rm max} = 3000,1500,500$ in red (filled contours, solid), green (unfilled contours, dashed) and blue (filled contours, dotted). Modelling of $C_\ell$ uses {\tt ReACT} + {\tt HMCode} (see equation~\ref{eq:nlps}). The mock data fiducial cosmological values are marked as dashed lines.}
\label{figuredgpellmax}
\end{figure*}

Similarly, figure~\ref{figuredgpellmax} shows the same results in DGP. Unlike the $f(R)$ case, we find a $25\%$ improvement in the marginalised constraints on the DGP parameter $\Omega_{rc}$ when including the multipoles between $\ell_{\rm max} = 3000$ and $1500$. Again, the constraints are severely degraded when using only quasi non-linear scales, with $\ell_{\rm max} = 500$.

Note for all analyses, we safely recover all fiducial cosmological parameter values (including $f_{\rm R0} \approx \Omega_{\rm rc} \approx 0$).


\begin{table}
\centering
\caption{Maximum multipole used in the analysis and the upper $68 (99.7) \%$ confidence bound on theory parameters around the marginalised mean values.}
\begin{tabular}{| c | c | c | }
\hline  
 \multicolumn{1}{ | c | }{$\ell_{\rm max}$} & ${\rm Log}[f_{R0}] + 1(3)\sigma$ & $\Omega_{rc} + 1(3)\sigma$ \\
 \hline
 3000 & -7.29 (-5.98) & 0.06 (0.28) \\ \hline 
 1500 & -7.25 (-5.86) & 0.08 (0.36) \\ \hline 
 500 & -6.96 (-5.46) & 0.42 (1.46) \\ \hline 
\end{tabular}
\label{fittable}
\end{table}

\subsection{$f(R)$ mock data}
Here we present the results for the F5 mock data. We remind the reader that these have the same Gaussian covariance matrix as the $\Lambda$CDM-mock data but the data vector is created using {\tt ReACT} + {\tt HMCode} (see equation~\ref{eq:nlps}). We fit this data both using $P^{\rm pseudo}_{\rm NL}(k)$ ($\mathcal{R}=1$) as modelled by {\tt HMCode} as well as the fully non-linear modelling including  $\mathcal{R}$. The motivation here is to inspect the impact of $\mathcal{R}$ on parameter estimation as well as estimate the capability of detecting a signal of $f(R)$ by including very non-linear scales.

\begin{figure*}
\centering
  \includegraphics[width=\textwidth,height=\textwidth]{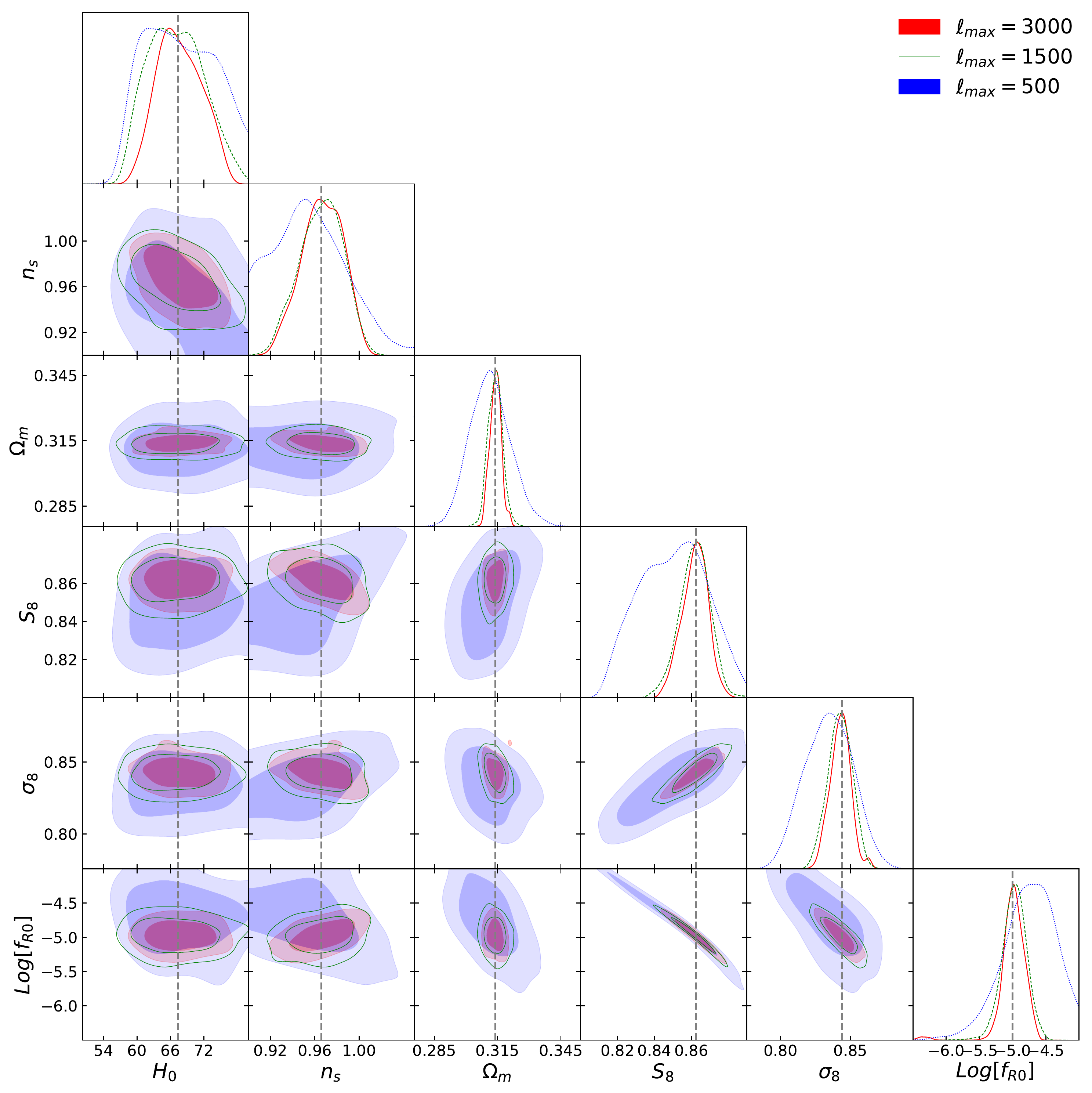}  \\ 
  \caption[CONVERGENCE]{MCMC results for an $f(R)$ analysis of an F5 mock data vector for an LSST-like survey, using $\ell_{\rm max} = 3000,1500,500$ in red (filled contours, solid), green (unfilled contours, dashed) and blue (filled contours, dotted). Modelling of $C_\ell$ uses {\tt ReACT} + {\tt HMCode} (see equation~\ref{eq:nlps}). The mock data fiducial cosmological values are marked as dashed lines.}
\label{figuref5}
\end{figure*}

Figure~\ref{figuref5} shows the marginalised posteriors for F5 for three analyses with different scale cuts: $\ell_{\rm max} = 3000,$ $\ell_{\rm max} = 1500$ and $\ell_{\rm max} = 500$, again using $19$, $15$ and $8$ $\ell$-bins respectively, using the fully non-linear model. We do not show the $\mathcal{R}=1$ model but report our results in table~\ref{fRfittable}.

We note that all analyses, including that using the $\mathcal{R}=1$ modelling, recover the fiducial value of $f_{\rm R0}$ within $1\sigma$. On the other hand, when the reaction is omitted from the modelling, strong degeneracies between various parameters appear, and biases are incurred, the largest being in $\sigma_8$ and $n_s$, as noted in table~\ref{fRfittable}. Further, setting $\mathcal{R}=1$ gives no clear detection of a modification to gravity, with very large errors around the marginalised mean being obtained. This indicates that the inclusion of $\mathcal{R}$ in the modelling is significantly important, in a stage IV survey context, for constraining $f_{\rm R0}$ at the $10^{-5}$-level.

We also note that when using the fully non-linear modelling, as well as including highly non-linear scales ($\ell_{\rm max}\geq 1500$), we make a clear detection of the non-zero value of $f_{\rm R0}$. If only linear to quasi non-linear scales are used then the constraints worsen by over a factor of 2. 

Finally, we have also done an analysis with an ${\rm Log}[f_{\rm R0}] = -6$ (F6) mock data vector. In this case the modification is much smaller and no biases are incurred on any parameters when using the $\mathcal{R}=1$ modelling. Further, the fiducial value of $f_{\rm R0}$ is also recovered safely within $68\%$ confidence using the $\mathcal{R}=1$. On the other hand, this choice of modelling again gives no clear detection of a non-zero value of $f_{\rm R0}$. 
\begin{table}
\centering
\caption{Maximum multipole used in the F5 mock data analysis and the upper and lower $68 \%$ confidence bound on selected parameters with the marginalised mean values. The fiducial cosmological values are $\sigma_8= 0.844$ and $n_s= 0.966$, with ${\rm Log}[f_{R0}]_{\rm fid} = -5$.}
\begin{tabular}{| c | c | c | c | }
\hline  
 \multicolumn{1}{ | c | }{$\ell_{\rm max}$} & $  {\rm Log}[f_{R0}] \pm 1 \sigma$ & $ \sigma_8 \pm 1 \sigma$ & $  n_s \pm 1 \sigma $  \\
 \hline
 3000 & $-5.00 \pm^{0.17}_{0.13}$ &  $0.843 \pm^{0.008}_{0.007}$  & $0.967 \pm^{0.021}_{0.017}$ \\ \hline
 1500 & $-4.98 \pm^{0.19}_{0.17}$ &  $0.842 \pm^{0.010}_{0.008}$  & $0.966 \pm^{0.022}_{0.018}$  \\ \hline
 500 & $-4.76 \pm^{0.45}_{0.32}$ &  $0.835 \pm^{0.018}_{0.017}$  & $0.956 \pm^{0.028}_{0.040}$ \\ \hline \hline
 3000 ($\mathcal{R}=1$) & $-5.80 \pm^{1.43}_{1.59}$ &  $0.861 \pm^{0.011}_{0.006}$  & $0.932 \pm^{0.008}_{0.032}$  \\ \hline 
\end{tabular}
\label{fRfittable}
\end{table}
\section{Summary}\label{sec:summary}
We have presented {\tt ReACT}, the first computational tool that accurately and reliably models the non-linear matter power spectrum in generic theories beyond $\Lambda$CDM employing the halo-model reaction introduced in \cite{Cataneo:2018cic} \footnote{Another approach to accurately model the non-linear power spectrum in theories beyond $\Lambda$CDM was recently suggested in \cite{Hu:2017aei,Ruan:2020irs}, called CHAM. We do not directly compare against CHAM but note that its level of accuracy is similar to {\tt ReACT} at $k\leq 1h/{\rm Mpc}$, yet is far more computationally expensive and so not currently suited for MCMC type analyses.}. We furthermore provide a {\tt CosmoSIS} implementation of this code that enables the performance of efficient MCMC parameter estimation analyses on the theory space. The merits of {\tt ReACT} and the halo-model reaction approach are summarised below:
\begin{itemize}
    \item 
    Framework to compute the non-linear power spectrum for wide class of theories beyond $\Lambda$CDM including scalar tensor theories and evolving dark energy without any fitting parameters.
    \item 
    Speed of computation allows for inference analyses in +6D parameter space on reasonable time scales. 
    \item
    Accuracy of code predictions for $f(R)$ and DGP are competitive with current state-of-the-art models/fitting formulae. 
    \item
    There is a clear direction for accuracy improvement: 
    \begin{enumerate}
        \item 
        Improving the accuracy of the pseudo $\Lambda$CDM power spectrum which is independent of {\tt ReACT}.
        \item
        Improving the halo model ingredients within {\tt ReACT}. This is easily implemented as all ingredients are stand alone functions within the code (see section~\ref{sec:code}).
    \end{enumerate}  
\end{itemize}
To demonstrate the performance of this tool, we have conducted various MCMC analyses for two test cases: $f(R)$ and DGP gravity. We employed $\Lambda$CDM mock data for these tests with a Gaussian covariance matrix based on LSST-like specifications. For these analyses, with an $\ell_{\rm max}=1500$, we find a forecasted constraint of $|f_{\rm R0}|\leq 10^{-5.86}$ for Hu-Sawicki $f(R)$ gravity and $\Omega_{rc}\leq 0.36$ for DGP at the $99.7\%$ confidence level.

Comparing our results to a similar analysis performed by \cite{Schneider:2019xpf,Schneider:2019snl} we find stronger constraints on $f_{\rm R0}$ by almost one order of magnitude. This discrepancy is readily explained by discrepancies in the modelling of $P_{\rm NL}$. In \cite{Schneider:2019xpf,Schneider:2019snl}, the authors use the fitting function of \cite{Winther:2019mus}. This fitting formula demonstrates relevant inaccuracies for values of $f_{\rm R0}$ not used in the fit. For example, for $f_{\rm R0} = 2\times10^{-5}$, the fitting function shows inaccuracies up to $5\%$  for $z\leq 2$ and $k\leq 10h/{\rm Mpc}$.

The reaction approach used here also demonstrates similar inaccuracies. In particular, the $P^{\rm pseudo}_{\rm NL}$ we adopt, which employs the halo model formula of \cite{Mead:2015yca}, is not trained for the exotic primordial power spectra used for the pseudo power spectrum, which will be remedied with the development of an emulator \citep{Giblin:2019iit}. As it stands, {\tt HMCode}  is $\sim 5\%$ accurate on scales $k\leq 10h/{\rm Mpc}$ in $\Lambda$CDM and is the dominant source of inaccuracy in the moderately non-linear regime ($0.1 < k < 3 h/{\rm Mpc}$). Above $k\sim3h/{\rm Mpc}$, the reaction prescription also becomes inaccurate and can be compounded with {\tt HMCode} inaccuracies. 

The inaccuracies in the modelling of the reaction, $\mathcal{R}(k,z)$, arise from its adoption of reinterpreted $\Lambda$CDM fitted halo-model ingredients such as the Sheth-Torman mass function. These inaccuracies are highlighted in figures~4 and~5 of \citet{Cataneo:2018cic}, which show that at $k\sim10 h/{\rm  Mpc}$ this basic version implementation, can have inaccuracies as large as $10\%$ for F5 and up to $3\%$ for F6, which are comparable to the inaccuracies of {\tt HMCode}. 

All this being said, we await ray tracing simulations of modified gravity in order to make more robust claims about the accuracy of our forecasts. We direct the reader to appendix~\ref{app:accuracy} for further discussion on the accuracy of the reaction approach used here and comparisons to the fitting function of \cite{Winther:2019mus}.

For surveys such as Euclid or LSST, the current pipeline can only be safely applied up to $\ell_{\rm max}\sim 500$.
Importantly, however, the current code is suitable to much higher $\ell_{\rm max}$ for current surveys such as KiDS \citep{Kuijken:2015vca}, HSC \citep{Aihara:2017paw} and DES \citep{Albrecht:2006um}, which have larger statistical uncertainties.
{\tt ReACT} is therefore readily applicable for parameter estimation analyses employing the currently available data.
We discuss further details in appendix \ref{app:accuracy}.  Note that work to improve the pseudo spectrum has already begun in \cite{Giblin:2019iit} and we expect to have a `pseudo-emulator' ready for stage IV survey lensing analyses.

We also investigate the impact of the reaction in separate sets of $f(R)$ mock data, created using {\tt ReACT} + {\tt HMCode}. We find that if the reaction is omitted from the modelling when fitting to this data, strong degeneracies and parameter biases may be incurred depending on the value of $f_{\rm R0}$, in particular between $n_s$, $\sigma_8$ and $f_{\rm R0}$. Thus, for consistent and unbiased constraints on cosmological parameters and gravity in the context of stage IV surveys, the reaction will be important. Further, if only scales within the quasi-linear regime are included, i.e. $\ell_{\rm max} \approx 500$, we find constraints on $f_{\rm R0}$ are degraded by over a factor of 2.

Next, we comment on model uncertainties coming from baryonic effects. Based on simulations and observations, these can be as large as $10-30\%$ for scales $1\leq k \leq 10h/{\rm Mpc}$ \citep{Chisari:2018prw,Schneider:2015wta,Semboloni:2011fe}. These are the scales at which the modelling discussed here becomes inaccurate, even though {\tt HMCode} provides a means to marginalise over this uncertainty. An emulator to address baryonic effects has been studied in \cite{Schneider:2019snl,Schneider:2019xpf} and a way to incorporate this emulator into this framework will be the goal of future work.

Further worth noting is that an extension of the current framework has also been developed to include massive neutrinos \citep{Cataneo:2019fjp,Wright:2019qhf} and its implementation into {\tt ReACT} will be the focus of an upcoming work. We also look to extend the current code to galaxy clustering observables, namely the redshift space power spectrum for biased tracers. This can be done by using the `hybrid' approach as outlined in \cite{Bose:2019yjp}, which  makes use of the Gaussian streaming model \citep{Reid:2011ar}. Work is also being conducted in upgrading the current implementation to include a general parametrisation of beyond-$\Lambda$CDM physics. Namely, we are looking to parametrise the 1-loop calculations in terms of the effective field theory of dark energy \citep[for example][]{Gubitosi:2012hu} using the reconstruction method described in \cite{Kennedy:2017sof,Kennedy:2018gtx}, and parametrising the spherical collapse using a PPF-formalism \citep[for example][]{Lombriser:2016zfz}. This would culminate in a tool that can comprehensively, consistently and accurately probe allowed deviations from $\Lambda$CDM at a very wide range of scales and for a very broad range of models.

\section*{Acknowledgments}
\noindent  BB and LL acknowledge support from the Swiss National Science Foundation (SNSF) Professorship grant No.~170547. CH, MC, TT, and QX acknowledge support from the European Research Council under grant number 647112. TT also acknowledges funding from the European Union’s Horizon 2020 research and innovation programme under the Marie Sk\l{}odowska-Curie grant agreement No.~797794. LL is also grateful for support from the Affiliate programme of the Higgs Centre for Theoretical Physics. CH acknowledges support from the Max Planck Society and the Alexander von Humboldt Foundation in the framework of the Max Planck-Humboldt Research Award endowed by the Federal Ministry of Education and Research. This work used the DiRAC@Durham facility managed by the Institute for Computational Cosmology on behalf of the STFC DiRAC HPC Facility (\url{www.dirac.ac.uk}). The equipment was funded by BEIS capital funding via STFC capital grants ST/P002293/1, ST/R002371/1 and ST/S002502/1, Durham University and STFC operations grant ST/R000832/1. DiRAC is part of the National e-Infrastructure 

\appendix

\section{Perturbation theory beyond $\Lambda$CDM }\label{app:modgpt}

For completeness we briefly describe here the perturbative approach to calculate a generalised density field, and hence the power spectrum at 1-loop order, used in section~\ref{sec:theory}. This is given by the expression
\begin{equation}
P_{\rm 1-loop}(k,a) = F_1^2(k,a)P_0(k) + P^{22}(k,a) + P^{13}(k,a),
\end{equation}
where $a$ is the scale factor, $P_0$ is the (linear) primordial power spectrum and $F_1(k,a)$ denotes the linear growth, which may be scale-dependent depending on the theory of choice. We begin with the density and  velocity divergence fields being treated perturbatively
\begin{equation}
\delta_{\rm NL}(\bfk,a) = \sum^{\infty}_{n=1}\delta_n(\bfk,a), \quad \quad \theta_{\rm NL}(\bfk,a) = \sum^{\infty}_{n=1}\theta_n(\bfk,a),
\end{equation}
 where
\begin{align}
  \delta_n(\boldsymbol{k},a) \sim & \int d^3\boldsymbol{k}_1...d^3 \boldsymbol{k}_n \delta_D(\boldsymbol{k}-\boldsymbol{k}_{1...n}) \nonumber \\ & \quad  \times F_n(\boldsymbol{k}_1,...,\boldsymbol{k}_n,a) \delta_0(\boldsymbol{k}_1)...\delta_0(\boldsymbol{k}_n) \label{densitypt}, \nonumber \\ 
    \theta_n(\boldsymbol{k},a) \sim & \int d^3\boldsymbol{k}_1...d^3 \boldsymbol{k}_n \delta_D(\boldsymbol{k}-\boldsymbol{k}_{1...n}) \nonumber \\ &  \quad \times  G_n(\boldsymbol{k}_1,...,\boldsymbol{k}_n,a) \delta_0(\boldsymbol{k}_1)...\delta_0(\boldsymbol{k}_n) \, ,
\end{align}
with $k_{1...n} = k_1 + k_2 ... + k_n$. The 1-loop matter power spectra terms are then given by 
\begin{align}
P^{22}(k,a) &= \int \frac{d^3p}{(2\pi)^3} F_2(\bfp,\bfk-\bfp,a)^2 P_0(p)P_0(|\bfk-\bfp|), \label{eq:p22} \\ 
P^{13}(k,a) &= 2F_1(k,a)P_0(k) \int \frac{d^3p}{(2\pi)^3} F_3(\bfp,-\bfp,\bfk,a) P_0(p). \label{eq:p13}
\end{align} 
To determine the kernels $F_i$, we solve the continuity and Euler equations order by order
 \begin{align}
& a  \delta'(\bfk)+\theta(\bfk) = \nonumber \\ & -
\int\frac{d^3\bfk_1d^3\bfk_2}{(2\pi)^3}\delta_{\rm D}(\bfk-\bfk_{12})
\alpha(\bfk_1,\bfk_2)\,\theta(\bfk_1)\delta(\bfk_2),
\label{eq:Perturb1}\\
& a \theta'(\bfk)+
\left(2+ \frac{a H'}{H}\right)\theta(\bfk)
-\left(\frac{k}{a\,H}\right)^2\,  \Phi(\bfk)= \nonumber \\ &
-\frac{1}{2}\int\frac{d^3\bfk_1d^3\bfk_2}{(2\pi)^3}
\delta_{\rm D}(\bfk-\bfk_{12})
\beta(\bfk_1,\bfk_2)\,\theta(\bfk_1)\theta(\bfk_2),
\label{eq:Perturb2}
\end{align}
where a prime denotes a derivative with respect to the scale factor and $\Phi$ is  the Newtonian   potential. The kernels $\alpha(\bfk_1,\bfk_2)$ and $\beta(\bfk_1,\bfk_2)$ are the standard mode coupling kernels 
\begin{align}
\alpha(\bfk_1,\bfk_2)= &1+\frac{\bfk_1\cdot\bfk_2}{|\bfk_1|^2}, \\
\beta(\bfk_1,\bfk_2)=&
\frac{(\bfk_1\cdot\bfk_2)\left|\bfk_1+\bfk_2\right|^2}{|\bfk_1|^2|\bfk_2|^2}.
\label{alphabeta}
\end{align}
Modifications to gravity enter through the Poisson equation
\begin{equation}
-\left(\frac{k}{a H(a)}\right)^2\Phi (\bfk,a)=
\frac{3 \Omega_{\rm m}(a)}{2} \mu(k,a) \,\delta(\bfk,a) + S(\bfk,a),
\label{eq:poisson1}
\end{equation}
where $\mu(k,a)$ is the linear modification to GR, while $S(\bfk,a)$ is a source term capturing non-linear modifications, including those responsible for screening effects. The source term is given by 
\begin{align}
 S(\bfk,a) & =
\int\frac{d^3\bfk_1d^3\bfk_2}{(2\pi)^3}\,
\delta_{\rm D}(\bfk-\bfk_{12}) \gamma_2(\bfk_1, \bfk_2,a)
\delta(\bfk_1)\,\delta(\bfk_2)
\nonumber\\
 & +
\int\frac{d^3\bfk_1d^3\bfk_2d^3\bfk_3}{(2\pi)^6}
 \delta_{\rm D}(\bfk-\bfk_{123})
\gamma_3(\bfk_1, \bfk_2, \bfk_3,a) \nonumber \\ & \quad \quad \times
\delta(\bfk_1)\,\delta(\bfk_2)\,\delta(\bfk_3). 
\label{eq:Perturb3}
\end{align}
The linear $\mu(k,a)$ and higher order $\gamma_i$ modifications to general relativity can be derived once we specify a particular theory. We refer the reader to appendix~\ref{app:parametrisations} for the forms of these functions in $f(R)$ and DGP. Furthermore, we note that this framework is very general and can encompass exotic dark energy models too \cite[see][]{Bose:2017jjx}. 

We can now calculate the $F_i$ kernels numerically by solving equations~\eqref{eq:Perturb1} and \eqref{eq:Perturb2} order by order, as described in \cite{Taruya:2016jdt,Bose:2016qun,Bose:2018zpk}, and so do not use the analytic forms which can be obtained by using the Einstein-de Sitter approximation as in \cite{Koyama:2009me}. The higher order kernels are then integrated as in equation~\eqref{eq:p22} and equation~\eqref{eq:p13} to calculate the 1-loop terms. 

\section{General spherical collapse} \label{app:sphercol}
In this work we use the Press-Schechter prescription~\citep{Press:1973iz}. In this approach, we trace the evolution of a spherical top-hat over-density $\delta$ with radius $R_{\rm TH}$ in a homogeneous background spacetime. The mass and momentum conservation equations govern the evolution of this over-density, and give the following second order differential equation for the top-hat radius \citep[e.g.][]{Schmidt:2008tn,Pace:2010sn}
\begin{equation}
\frac{\ddot{R}_{\rm TH}}{R_{\rm TH}} = -\frac{4\pi G}{3} [\bar{\rho}_m +(1+3w)\bar{\rho}_{\rm eff}] - \frac{1}{3}\nabla^2 \Phi. \label{eq:sphercol}
\end{equation}
 $\bar{\rho}_m$ is the background matter density, $\bar{\rho}_{\rm eff}$ is the background energy density of an effective dark energy component having equation of state $w$. In this work we assume a $\Lambda$CDM background for DGP and $f(R)$ theories and so in these cases, $\bar{\rho}_{\rm eff} = \bar{\rho}_\Lambda$, the energy density of the cosmological constant, with $w=-1$. For these gravity models, we have a modification to the Poisson equation 
\begin{equation}
\nabla^2 \Phi = 4\pi G (1+\mathcal{F})\bar{\rho}_{\rm m} \delta,
\label{eq:poisson2}
\end{equation}
where $\mathcal{F}$ parametrises the dependency on the theory of gravity with $\mathcal{F}=0$ for general relativity.

Using equation~\ref{eq:sphercol} we can derive certain key quantities used in  prescriptions for the halo mass function, virial concentration and halo density profile. These are then used to compute the one-halo term $P_{\rm 1h}(k,z)$ that goes into the reaction, $\mathcal{R}(k,z)$ (see equation~\eqref{eq:reaction}).

One such key quantity is the non-linear over-density,  which can be written as a function of top hat radius as 
\begin{equation}
    \delta = \left( \frac{R_i}{R_{\rm TH}(a)} \right)^3 (1+\delta_i) -1, \label{eq:nldensity}
\end{equation}
where a subscript `i' denotes the initial value of the quantity. Given a scale factor of collapse, $a_{\rm col}$, we seek the corresponding initial over-density $\delta_i$ that will provide a collapsed object at that time, i.e. $R_{\rm TH}(a_{\rm col}) = 0$. Once we have this value of $\delta_i$ one can extrapolate to the desired time using linear theory i.e. $\delta_c(a) = D(a) \delta_i / a_i$ where $D(a)$ is the linear growth rate (see equation~\ref{densitypt}) in $\Lambda$CDM.

As over-densities collapse, they begin to virialise and eventually become stable structures known as halos. To determine the time when such halos form ($a_{\rm vir}$), we can solve the virial theorem  which can include modified gravity or dark energy (see equations A6-A11 of \cite{Cataneo:2018cic} for example). The time of virialisation can  be used to get another key quantity, the over-density at the time of virialisation \citep[e.g.][]{Schmidt:2008tn}
\begin{equation}
    \Delta_{\rm vir} = [1+\delta(a_{\rm vir})] \left(\frac{a_{\rm col}}{a_{\rm vir}} \right)^3.
    \label{deltavir}
\end{equation}
The corresponding mass of this spherical halo is then
\begin{equation}
    M_{\rm vir} = \frac{4\pi}{3} R_{\rm vir}^3 \bar{\rho}_{m,0}\Delta_{\rm vir},
\end{equation}
where $\bar{\rho}_{m,0}$ is the background matter density today and $R_{\rm vir}$ is the comoving radius of this halo.

\section{Model parametrisations} \label{app:parametrisations}

Next we give explicit forms for the parameters currently implemented in {\tt ReACT}. These include a function characterising the modification to spherical collapse, $\mathcal{F}$, and three functions  characterising the modification to each order in perturbation theory up to 3rd order, required for the 1-loop power spectrum computation. Additionally, the background expansion needs to be specified. For all modified gravity scenarios considered in this  work, we assume a $\Lambda$CDM expansion history

\begin{equation}
    \frac{H(a)}{H_0} = \sqrt{\frac{\Omega_{m,0}}{a^3} + \Omega_\Lambda},
\end{equation}

\noindent where $\Omega_{m,0}$ is the total matter energy density fraction today and $\Omega_\Lambda$ is the cosmological constant energy density fraction. In our case we have $\Omega_\Lambda = 1 - \Omega_{m,0}$. $H_0$ is the Hubble constant.

We also write here the explicit form of the spherical collapse equation coded up in {\tt ReACT}
\begin{equation}
    y'' + \frac{H'}{H} y'  - \left(1+\frac{ H'}{H}\right) y + \left[\frac{H_0}{H} \right]^2 \frac{\Omega_{m,0} }{2 a^3} (1 + \mathcal{F}) \delta (\frac{a}{a_i} + y) = 0,  \label{eq:sphercol2}
\end{equation}
where $y = R_{\rm TH}/R_i - a/a_i$ and the prime denotes derivatives $d/d\ln a$, and $a_i$ is the initial scale factor. 

We have considered two modifications to gravity to test {\tt ReACT}: Hu-Sawicki $f(R)$ gravity \citep{Hu:2007nk} and the normal branch of DGP \citep[][]{Dvali:2000hr}. Note that in the current form, the code requires the specification of the theory parameters for the particular theory under consideration. A future goal is to implement a generalised parametrisation for modified gravity and dark energy.

\subsection{DGP gravity}
\label{subsubsec:nDGP}
 The DGP model~\citep{Dvali:2000hr} of gravity assumes that we live on a four-dimensional brane embedded in a five-dimensional spacetime bulk. The linear modification to the Poisson equation, $\mu(k,a)$, is given in the normal branch of DGP  by 
\begin{equation}
\mu(k,a) = 1 + \frac{1}{3\beta}  \ , \qquad 
 \label{betadef}
\beta(a) \equiv 1+\frac{H}{H_0} \frac{1}{\sqrt{\Omega_{\rm rc}}}   \left(1+\frac{aH'}{3H}\right) \,.
\end{equation}
Here we choose to parameterise the additional free parameter of the theory, the scale at which gravity dilutes into the 5th dimension, $r_c$, in terms of the associated current fractional energy density $\Omega_{\rm rc} \equiv 1/(4r_{\rm c}^2H_0^2)$. Current cosmological constraints limit the crossover distance $r_{\rm c}$ to a few times the Hubble length~\citep{Lombriser:2009xg,Raccanelli:2012gt,Barreira:2016ovx}. The higher order coupling kernels are given by \citep{Bose:2016qun}
\begin{equation}
\gamma_2(\bfk_1,\bfk_2,a) = -\left[\frac{H_0}{H}\right]^2\frac{1}{24\beta(a)^3 \Omega_{\rm rc}} \left(\frac{\Omega_{\rm m0}}{a^3}\right)^2 (1-\mu_{1,2}^2) ,
\end{equation}
\begin{align}
\gamma_3(\bfk_1,\bfk_2,\bfk_3,a) = & \left[\frac{H_0}{H}\right]^2\frac{1}{144 \beta(a)^5 \Omega_{\rm rc}^2} \left(\frac{\Omega_{\rm m0}}{a^3}\right)^3 \nonumber \\ & \times (1-\mu_{2,3}^2) (1-\mu_{1,23}^2),
\end{align}
where $\mu_{i,j} = \hat{\bfk}_i\cdot \hat{\bfk}_j$ is the cosine of the angle between $\bfk_i$ and $\bfk_j$ and  $\boldsymbol{k}_{1...n} = \boldsymbol{k}_1 + ...+ \boldsymbol{k}_n$. 

Finally, the spherical collapse modification takes the form \citep{Schmidt:2009yj}
\begin{equation}
    \mathcal{F} = \frac{2}{3\beta(a)} \frac{\sqrt{1+s^{3}} -1}{s^{3}},
\end{equation}
where 
\begin{equation}
    s  = \left[ \frac{2 \Omega_{m,0} \delta }{9 a^3 \beta(a)^2 \Omega_{rc}} \right]^{\frac{1}{3}},
\end{equation}
$\delta$ being the non-linear over-density given in equation \eqref{eq:nldensity}.

\subsection{Hu-Sawicki $f(R)$ gravity}
\label{subsubsec:fR_gravity}

$f(R)$ gravity is a class of models in which the Einstein-Hilbert action is generalised to include an arbitrary non-linear function of the scalar curvature. Among various examples for the functional form of $f(R)$, the \cite{Hu:2007nk} model is particularly well studied  \citep{Song:2015oza,Hammami:2015iwa,Lombriser:2013wta,Hellwing2013,Zhao:2013dza,Okada:2012mn,Li:2011pj,Lombriser:2011zw,Lombriser:2010mp,Schmidt:2009am,Brax:2008hh,Song:2007da,Burrage:2017qrf,Cataneo:2016iav,Cataneo:2014kaa} and provides a simple form with which chameleon-type screening is realised. It is given by
\begin{equation}
f(R)  = -m^2 \frac{c_1 (R/m^2)^n}{c_2(R/m^2)^n+1}.
\label{husawicki}
\end{equation}
Current cosmological constraints on this parameter lie in the range of $|f_{\rm R0}|\lesssim\left(10^{-6}-10^{-5}\right)$ (see table~1 of \cite{Lombriser:2014dua} for a summary).
In this work, we specifically consider the $n=1$ case. Then, the $f(R)$ form of the Poisson equation  \citep[see for example][]{Koyama:2009me,Taruya:2014faa}, is characterised by 
 \begin{align}
\mu(k,a) = &1 + \left(\frac{k}{a}\right)^2\frac{1}{3\Pi(k,a)}, 
\\ 
\gamma_2(\bfk_1,\bfk_2,a)  = &- \frac{3}{16}\left(\frac{kH_0}{aH}\right)^2\left(\frac{\Omega_{\rm m,0}}{a^3}\right)^2  \frac{\Xi(a)^5}{f_0^2 (3\Omega_{\rm m,0}-4)^4} \nonumber \\ & 
\times \frac{1}{\Pi(k,a)\Pi(k_1,a)\Pi(k_2,a)}, 
\label{frg2} 
\end{align}
and
\begin{align}
&\gamma_3(\bfk_1,\bfk_2,\bfk_3,a)   = \nonumber \\ & \frac{1}{32} \left(\frac{kH_0}{aH}\right)^2   \left(\frac{\Omega_{\rm m,0}}{a^3}\right)^3 \frac{1}{\Pi(k,a)\Pi(k_1,a)\Pi(k_2,a)\Pi(k_3,a)}  
\nonumber 
\\ & \times   \left[-5\frac{\Xi(a)^7}{f_0^3(3\Omega_{\rm m,0}-4)^6}  + \frac{9}{2}\frac{1}{\Pi(k_{23},a) }\left( \frac{\Xi(a)^5}{ f_0^2 (3\Omega_{\rm m,0}-4)^4} \right)^2\right],
\label{frg3} 
\end{align}
where the functions $\Pi$ and $\Theta$ are given by
\begin{align}
\Pi(k,a) =& \left(\frac{k}{a}\right)^2+\frac{\Xi(a)^3}{2f_0(3\Omega_{\rm m,0}-4)^2}, \\   \Xi(a) =& \frac{\Omega_{\rm m,0}+4a^3(1-\Omega_{\rm m,0})}{ a^3},
\end{align}
with $f_0 = |{f}_{\rm R0}|/H_0^2$. $f_{\rm R0}$ is the current amplitude of $f_R = df(R)/dR$, which can be interpreted as the current amplitude of the scalar field in an equivalent scalar-tensor-theory (e.g. present scalaron amplitude). 

Finally, the spherical collapse modification takes the form \citep{Lombriser:2013eza}

\begin{equation}
    \mathcal{F} = {\rm min} \left[O - O^2 + \frac{O^3}{3}  , \frac{1}{3} \right] ,
\end{equation}
where 
\begin{equation}
    O = \frac{ f_0 x_h a (3 \Omega_{m,0} - 4)^2}{ \Omega_{m,0} R_{\rm TH}^2} \times
    \left[ \tilde{G}(x_{\rm env}) -  \tilde{G}(x_{\rm h}) \right] \, ,
\end{equation}
and 
\begin{equation}
    \tilde{G}(x) = \left[ \frac{\Omega_{m,0}}{(x a)^3} +4 - 4\Omega_{m,0} \right]^{-2} \, , 
\end{equation}
with 
\begin{equation}
    x = \frac{R_{\rm TH}}{R_i} \frac{a_i}{a} ,
\end{equation}
$x_h$ being the quantity solved for the $f(R)$ halos whereas $x_{\rm env}$ is that for the environment, i.e. with $f_0 = 0$.

\section{Accuracy of {\tt ReACT}}\label{app:accuracy}
The halo model reaction approach and {\tt ReACT} have been shown to be very promising means of modelling the matter power spectrum, with {\tt ReACT} able to be applied to general beyond-$\Lambda$CDM models, requiring no fits, nor simulation measurements. Here we aim to detail the current code version's accuracy, that is the reaction combined with a pseudo power spectrum as predicted by {\tt HMCode}. As a test case, we will consider $f(R)$  gravity. Means of improving the overall accuracy of {\tt ReACT} are clear and are discussed briefly in section~\ref{sec:summary}. Such improvements would be applicable to all models of gravity and dark energy and not just $f(R)$, such as the development of an emulator for the pseudo power spectrum \cite{Giblin:2019iit}.

\begin{figure*}
\centering
  \includegraphics[width=\textwidth,height=10cm]{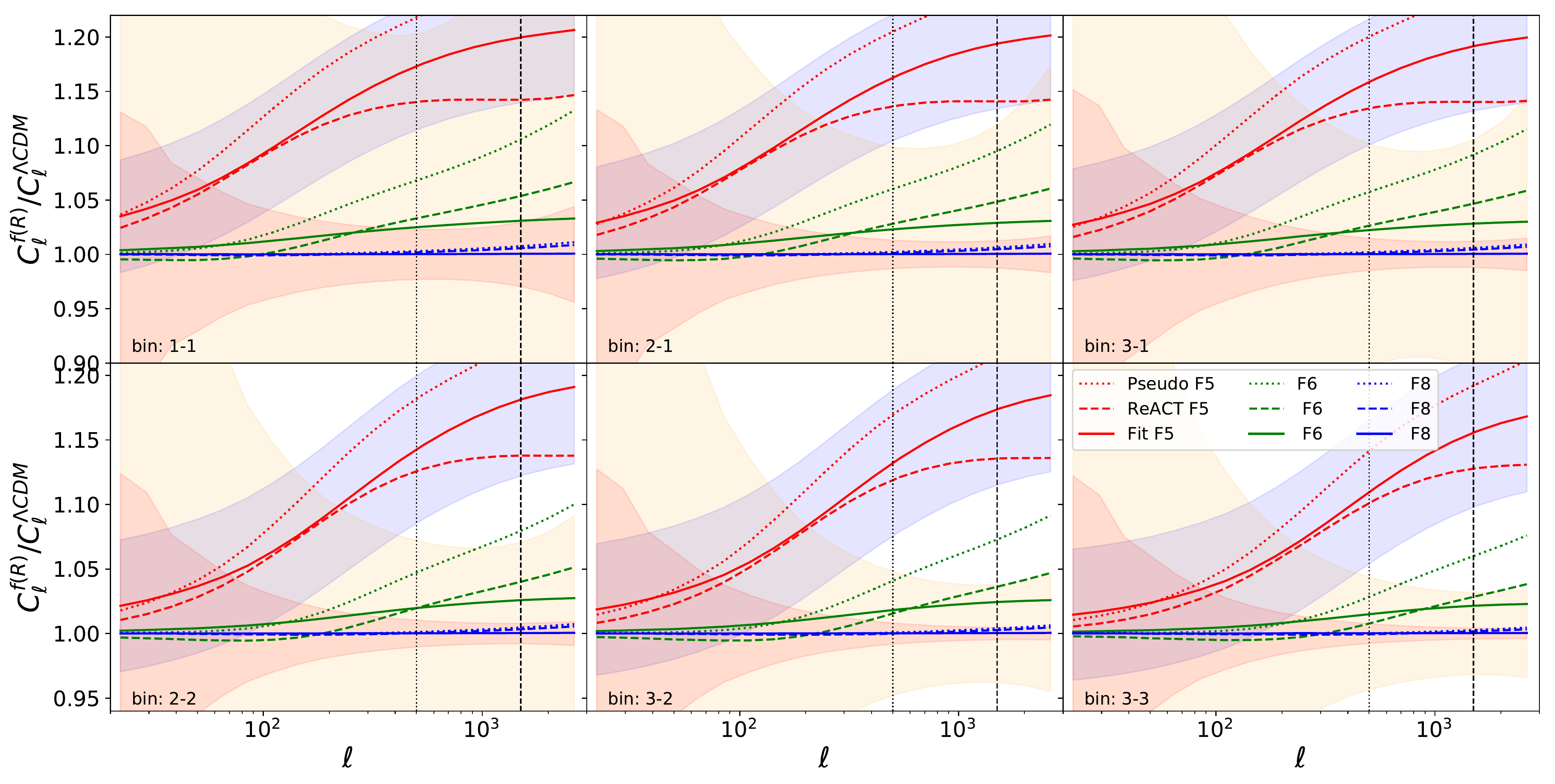}  \\  
  \caption[CONVERGENCE]{Ratio of cross and auto correlations between 3 tomographic bins in the redshift range of $z\in[0,2]$ of the cosmic shear $C_\ell$ between $f(R)$ gravity and $\Lambda$CDM. Predictions for the pseudo spectrum using {\tt HMCode} are shown as dotted lines, the reaction-corrected pseudo spectrum using {\tt ReACT} and {\tt HMCode} are shown as dashed lines and the fitting formula of \cite{Winther:2019mus} are shown as solid lines. Red curves correspond to a value of $\log[f_{\rm R0}] =-5 $, green curves correspond to a value of $\log[f_{\rm R0}] =-6 $, whereas blue curves correspond to the value  $\log[f_{\rm R0}] =-8.0$. The beige $1 \sigma$ Gaussian error bands represent a KiDS-like survey while the red error bands represent an LSST-like survey. The blue band represents a $5\%$ systematic uncertainty in the modelling. The dashed line predictions took a total of 59 seconds to obtain on a single core of the baobab cluster at the University of Geneva. 
  }
\label{figurecl2}
\end{figure*}

Recently, an accurate fitting formula for the Hu-Sawicki model was developed in \cite{Winther:2019mus}, which is also the formula adopted in the cosmic shear mock data analysis of \cite{Schneider:2019xpf}, where the authors find constraints of $\log [f_{\rm R0}] < -5.7$ at $1\sigma$ significance for an LSST-like survey. In this work, using similar survey specifications and $\ell_{\rm max}$ we find constraints that are an order of magnitude more stringent. This is largely related to an excess of power predicted by {\tt ReACT} + {\tt HMCode} over the fitting formula for $\log[f_{\rm R0}]\leq -6$ (see  figure~\ref{figurecl2}). Further, differences in the analyses, such as mock data uncertainties, linear theory and treatment of systematics such as intrinsic alignments will also play a role. 

This being said, the {\tt ReACT} + {\tt HMCode} pipeline applied in this work and the fitting formula share inaccuracies of a similar order at scales $k\leq 1h/{\rm Mpc}$ ($\sim 2\%$)  and $1 \leq k \leq 5 h/{\rm Mpc}$ ($\sim 5\%$). Above $k\approx 5h/{\rm Mpc}$, inaccuracies in the reaction become non-negligible and when compounded with the inaccuracies from {\tt HMCode} make {\tt ReACT} + {\tt HMCode} less accurate than the fitting formula. The limitations of the reaction approach used in this work were already noted in \cite{Cataneo:2018cic} (see figure 6 of this reference, for example). One can see the limitations of the fitting formula approach in figure 6 and figure 7 of \cite{Winther:2019mus}. 

In figure~\ref{figurecl2} we show the ratio of the auto and cross cosmic shear $C_\ell$ in $f(R)$ gravity to the same quantity in $\Lambda$CDM over all tomographic bins used in the analyses of section~\ref{sec:analysis}. We show the predictions as given by {\tt ReACT} + {\tt HMCode} and  the fitting formula of \cite{Winther:2019mus}. We also show the {\tt HMCode} prediction for the pseudo power spectrum for comparison. The $\Lambda$CDM spectrum in all these cases are computed by setting $f_{R0} = 0$ in the respective codes. 

Finally, overlaid are representative $1 \sigma$ Gaussian error bands of two different surveys: a future experiment such as LSST (same errors as in section~\ref{sec:analysis}) and an ongoing/finished one such as KiDS \citep{Kuijken:2015vca} for which we assume a sky fraction of $f_{\rm sky} = 0.024$,  number density of galaxies per arcminute squared $n = 3.5 {\rm arcmin}^{-2}$ per tomographic bin and shape noise parameter $\sigma_e = 0.28$. The tomographic binning in both cases is as follows. Bin $1: [0.00, 0.48]$, bin $2: [ 0.48, 0.81]$ and bin $3:[0.81, 2.00]$.  We also show a $5\%$ error band around the fitting formula prediction for F5 (red solid), which represents systematic errors coming from modelling uncertainty. This indicates that the differences between the modelling used here and that of \cite{Winther:2019mus} are consistent to within the systematic uncertainty. This result is also confirmed in \cite{Bose:2019yjp} where the authors find the fitting formula and the reaction approach to be consistent within $5\%$ at scales $k\leq 3h/{\rm Mpc}$ and for $z\leq 1$.

We find that for a KiDS-like survey, both the reaction and fitting formula modelling for $C_\ell$ are acceptable up to $\ell_{\rm max} \sim 3000$ if we want to constrain $\log[f_{\rm R0}]$ down to the $-6$-level, which is roughly the limit of this survey given our specifications. Note that this of course assumes statistical errors only and accounting for systematics will significantly degrade this upper bound. For tighter constraints than this, modelling needs to be improved as the uncertainty between these two state-of-the-art approaches becomes comparable to the statistical uncertainty of the measurement. 

For an LSST-like survey, the differences between the two approaches becomes far larger than the statistical uncertainty at small scales (large $\ell$). {\tt HMCode} is not sufficiently accurate for such a survey and so improvement needs to be made here before this methodology can be safely applied to these surveys. Such improvement is already underway \citep{Giblin:2019iit}.

\bibliographystyle{mnras}
\bibliography{mybib} 

\bsp	
\label{lastpage}
\end{document}